Resistivity minimum in strongly phase separated manganite thin films: impact of intrinsic and extrinsic perturbations

Vasudha Agarwal, Geetanjali Sharma, P. K. Siwach, K. K. Maurya and H. K. Singh[#]
CSIR-National Physical Laboratory, Dr. K. S. Krishnan Marg, New Delhi-110012, India

Abstract

The origin of the resistivity minimum observed in strongly phase separated manganites has been investigated in single crystalline thin films of $La_{1-x-y}Pr_yCa_xMnO_3$ ($x \approx 0.42$, $y \approx 0.40$). The antiferromagnetic/charge ordered insulator (AFM/COI)-ferromagnetic metal (FMM) phase transition, coupled with the colossal hysteresis between the field cool cooled and field cooled warming magnetization demonstrates strongly phase separated nature, which gives rise to non-equilibrium magnetic liquid state that freezes into a magnetic glass. The thermal cycling and magnetic field dependence of the resistivity unambiguously shows that the pronounced resistivity minimum observed during warming is a consequence non-equilibrium states resulting from the magnetic frustration created by the delicate coexistence of the FMM and AFM/COI phases. The non-equilibrium states and hence the resistivity minimum is extremely sensitive to the relative fraction of the coexisting phases and can be tuned by intrinsic and extrinsic perturbations like the defect density, thermal cycling and magnetic field.

[#] Corresponding Author; Email: hks65@nplindia.org

In doped rare earth manganese oxides of type $RE_{1-x}AE_xMnO_3$ (RE=La, Nd, Pr, Sm, etc. and AE=Ca, Sr, Ba, etc.), popularly called manganites, which have been recognised as prototype strongly correlated magnetic systems, the coupled charge, spin, orbital and lattice degrees of freedom give rise to variety of electronic phases like paramagnetic insulator (PMI), ferromagnetic metal (FMM), antiferromagnetic charge ordered insulator (AFM/COI), etc.[1,2] Mechanisms like double exchange (DE), superexchange (SE), electron-lattice coupling through the Jahn-Teller (JT) distortion of the $MnO_6$ octahedra, etc. invoked to explain the observed properties and phase transitions have met with restricted success.[1-4] The relative strength of the magnetic interactions governing these phases and the nature of transition depends on the one electron bandwidth (BW) of the system, which is determined by the average size of the RE/AE site cations and their size difference.[2] At reduced BW, the small difference between the ground state energies of coexisting electronic phases enhances the competition between the coexisting phases, mainly FMM and AFM/COI and leads to mesoscopic scale phase separation (PS). Of late PS has been recognized as the unavoidable and key ingredient of the physics of manganites.[2,5] In this context $Sm_{1-x}Sr_xMnO_3$ (0.4≤x≤0.55) and $La_{1-x-y}Pr_yCa_xMnO_3$ (x>0.3, y>0.4) have emerged as the representative mesoscopic PS.[6-15] The balance between FM and AFM phases gives rise to magnetic frustration, which in turn leads to non-equilibrium states below the FM transition ($T_C$), e.g., the magnetically disordered liquid that freezes into a random strain glass (SRG)[10-12] below the glass transition temperature ($T_g$). The liquid like regime is characterized by several features representative of first order phase transition; like (i) strong hysteresis between the field cooled cool (FCC) and field cooled warming (FCW) magnetization (M-T), (ii) hysteretic behaviour of the temperature dependent resistivity (ρ-T) measured in cooling and warming cycles, and (iii) sharp drop ρ (T) at moderate magnetic field. The frozen state at $T< T_g$ is characterized by thermal reversibility of M-T and ρ-T. All the above mentioned characteristics have been observed in the prototypical phase separated system $La_{1-x-y}Pr_yCa_xMnO_3$ for wide range of x and y.[9-17]

In strongly phase separated manganites like $La_{1-x-y}Pr_yCa_xMnO_3$, the drastic difference between the ρ-T behaviour during cooling and warming cycles is demonstrated by lower value of the $T_C$ and the insulator-metal transition (IMT) temperature ($T_{IM}$) during the former. During the warming cycle $T_C/T_{IM}$ is enhanced with a concomitant decrease in ρ (T). Another important feature observed in the warming cycle is the minimum ($\rho_m (T_m)$) in the ρ-T curve at $T=T_m>T_g$. Sathe et al.[16] have observed such minimum in 200 nm thick $La_{5/8-y}Pr_yCa_{3/8}MnO_3$ (y=0.45) film on SrTiO$_3$ (100) substrates and have attributed the $\rho_m$ (T$_m$) to melting of the blocked glassy state and its subsequent crystallization into crystalline FMM.[17] Their results show that as the magnetic field (H) applied in the warming cycle was increased both T$_g$ and T$_m$ decreased and cooling and warming at the same H leads to the disappearance of $\rho_m$ (T$_m$), while warming in a higher H leads its reappearance. Recently, $\rho_m$ (T$_m$) has been observed in 140 nm thick La$_{1-x-y}$Pr$_y$Ca$_x$MnO$_3$ (x≈0.42, y≈0.40) films (~140 nm) on SrTiO$_3$ (110)[18] as well as 40 nm thin La$_{5/8-y}$Pr$_y$Ca$_{3/8}$MnO$_3$ (y≈0.4) films on SrTiO$_3$ (001)[19]. These results clearly show that the resistivity minimum in low BW phase separated manganites is observed over a wide range of compositions and film thickness and hence it could not be related to perturbations like substrate induced strain or nanostructuring. Rawat et al.[20] have observed $\rho_m$ (T$_m$) in nanocrystalline Pr$_{0.7}$Ca$_{0.3}$MnO$_3$ even when the sample was cooled and warmed in equal fields. This suggests that $\rho_m$ (T$_m$) is related to the relative fraction of AFM/COI and FMM phases.

The results summarized above suggest that $\rho_m$ (T$_m$) at T$_m$>T$_g$ is the function of the relative fraction of the coexisting FMM and AFM/COI and caused by dominance of the later over former. However, the impact the substrate induced strain and its relaxation (via formation of defects) on the nature of the non-equilibrium phases and the resistivity minimum $\rho_m$ (T$_m$) in strongly phase separated manganites like La$_{1-x-y}$Pr$_y$Ca$_x$MnO$_3$ has not been studied. In present work we have investigated the tunability of the non-equilibrium phases and the resistivity minimum in La$_{1-x-y}$Pr$_y$Ca$_x$MnO$_3$ (x≈0.42, y≈0.40) in terms of intrinsic and extrinsic perturbations like defect density, magnetic field and thermal cycling. Our results show that the ρ-T minimum is a consequence of the non-equilibrium nature of the magnetic liquid and the frozen magnetic frustration and it disappears when the system is pushed towards the equilibrium by tilting the magnetic balance in favour of FMM phase.

The La$_{1-x-y}$Pr$_y$Ca$_x$MnO$_3$ (x≈0.42, y≈0.40) (thickness ~100 nm) thin films were grown by RF magnetron sputtering of two inch diameter target in 200 mtorr of Ar + O$_2$ (80 % + 20 %) mixture on single crystal LaAlO$_3$ [LAO, (001)] substrate (5x3x0.5 mm$^3$) maintained at temperature ~ 800 °C. After deposition the films were annealed insitu at 850 °C for 4 hrs in ~360 torr oxygen pressure. To test the reproducibility of the reults we prepared three set of films under identical conditions. The strcutructural and microstructural characteristics were probed by high resolution X-ray diffraction (HRXRD, PANalytical PRO X'PERT MRD, Cu-Kα1 radiation λ = 1.5406 Å) and atomic force microscopy (AFM, VEECO Nanoscope VI), respectively. The magnetic and magnetotransport properties were measured by commercial magnetic property measurement system (MPMS, Quantum design) and physical property measurement system (PPMS, Quantum Design), respectively.

2θ/ω scan presented in the Fig.1 shows fully oriented growth along (00ℓ). The value of out of plane lattice constant estimated from the experimental data is found to be larger than that of the target material ($a_c^{film}$ = 3.844 Å > $a_c^{bulk}$ = 3.822 Å). The average value of the in plane lattice parameter calculated from the (110) asymmetric 2θ/ω scan data is $a_{in}$ = 3.831 Å, which is very close to the bulk value. The fact that $a_c^{film}$ > $a_c^{bulk}$ clearly indicates the presence of small degree of compressive strain. The substrate induced strain shows relaxation tendencies at film thickness higher than 60-70 nm[21] and hence at film thickness ~100 nm, partial relaxation of the strain is expected. However, in view of the delicate balance between the coexisting electronic phases even a small degree of strain can tilt the balance and have explicit impact on the magnetotransport properties. The very small full width at half maximum (FWHM) Δω≈0.09° of the ω-scan of the (002) diffraction peak presented in the inset of Fig.1 is indicative of high quality of the films.

The temperature dependence of magnetization (M-T) measured employing zero field cooled (ZFC), field cooled cooling (FCC) and field cooled warming (FCW) protocols is shown in Fig. 2. The transition temperatures were determined from the first order derivative of the M-T data shown in the inset. The absence of explicit CO transition around $T_{CO}$ ≈ 230 K suggests quenching of the COI phase. M-T curves show broad shoulder like feature in the temperature range 185 K ≤T≤ 134 K, which could be due to the AFM ordering with Neel temperature ($T_N$) ≈ 167 K. However, nature of the broadening suggests the presence of COI alongside the AFM. On further lowering the temperature, FM ordering appears, which is evidenced by the rise in M-T around ≈134 K. The appearance of FM correlations is manifested differently in the three M-T curves. ZFC, FCC and FCW data of yield three distinct $T_C$ values for the three measurement protocols; $T_C^{ZFC}$ ≈ 123 K, $T_C^{FCC}$ ≈ 52 K and $T_C^{FCW}$ ≈ 125 K. At T ≤ $T_C$, ZFC and FCW curves show pronounced divergence, generally attributed to a metamagnetic state akin to cluster glass.[21-23] Further, the FCC-FCW curves show colossal hysteresis, which along with the large difference between the $T_C^{FCC}$ and $T_C^{ZFC}$/$T_C^{FCW}$ is due to the supercooling of the magnetic liquid.[10-16,18,19] The magnetic frustration due to competitive coexistence of FM and AFM-COI phases hinders the crystallization of FM order at the equilibrium $T_C$. The sharp decline in the ZFC at $T_P$ ≈ 52 K and the isothermal peak in the FCW curve could be regarded as the onset of the freezing of the magnetic liquid phase. The slope change at $T_g$ ≈ 28 K could be regarded as the glass transition point.[11] This also supported by the thermal reversibility and weak temperature dependence of the FCC-FCW curves at $T<T_g$. The $\rho$-T data measured in cooling and warming cycles is plotted in Fig. 3. Just like the FCC-FCW M-T the zero field $\rho$-T also shows a mega hysteresis with respect to thermal cycling. In the cooling cycle the IMT occurs at $T^C_{IM} \approx 56$ K where the resistivity drops by more than five orders of magnitude. Further, similar to the FCC-FCW M-T data the $\rho$-T curves also exhibit saturation like behaviour at $T<25$ K. In the warming cycle $\rho$-T remains reversible till about $T\approx 25$ K and then decreases sharply approaching a minimum ($\rho_m$) at $T_m\approx 56$ K. At $T>T_m$ $\rho$ increases and IMT is observed at $T^W_{IM} \approx 131$ K. Thus the hallmark of the zero field $\rho$-T is the appearance of colossal hysteresis with $\Delta T_{IM} = T^W_{IM} - T^C_{IM} = 75$ K and the resistivity minimum at $T_m$.

The M-T and $\rho$-T results presented above show that the colossal thermal hysteresis observed in the FCC-FCW M-T ($\Delta T_C = T^W_C - T^C_C = 73$ K) and zero field cooled and warmed $\rho$-T ($\Delta T_{IM} = T^W_{IM} - T^C_{IM} = 75$ K) is significantly stronger than the generally reported in thin films.[12-16,18-20] We believe that such strong hysteresis apart from being signature of the first order FMM-AFM/COI phase transition, is also a manifestation of the strongly frustrated magnetic state created by the delicate balance between two differently ordered magnetic phases. The observed reversibility of M-T and $\rho$-T at $T<T_g$ is a manifestation of the transition of the non-equilibrium magnetic liquid to another such phase, which in this case is randomly frozen glassy state akin to SRG. The non-equilibrium nature of the SRG is evident from the FCW M-T maximum at $T_p\approx 52$ K and $\rho$-T minimum at $T_m\approx 56$ K during the warming cycle. The fact that the resistivity at $T_m$ is found to be even less than half of its value at $T<T_g$ could be regarded as the measure of the strength of the randomness and the degree of freezing in the SRG regime. During the warming cycle the devitrification of the non-equilibrium glassy state formed by the frozen magnetic liquid gives rise to a thermally activated recrystallization into an equilibrium FMM phase. At this point we would like emphasize that the $\rho_m (T_m)$ observed in the present case is drastically different from the resistivity minimum observed in intermediate and large BW manganites.[24,25]

The nature of the non-equilibrium states, e.g., the magnetic liquid and glassy state is believed to be sensitive to the relative fraction of the FMM and AFM/COI phases.[16-20] If the observed resistivity minimum (($\rho_m(T_m)$)) is a manifestation of the non-equilibrium SRG to equilibrium FMM phase transition, then is must be a function of the relative fraction of the two coexisting phases, viz., FMM and AFM/COI. If the fraction of the FMM phase is enhanced then $\rho_m(T_m)$ should either dilute or vanish. The FMM fraction can be enhanced by intrinsic as well as extrinsic perturbations, such as the defect density, magnetic field (H) and selective thermal cycling as proposed by Singh et al.[19] Singh et al.[19] have shown that if a film showing colossal ρ-T hysteresis is warmed up from the SRG state and then cooled down from a temperature $T^*<T_{IM}$, the supercooling transition at $T_{IM}^C$ is shifted to a higher temperature but the $T_{IM}^W$ remains unchanged. Using similar protocol we cooled down the film under investigation from 128 K to 5 K and then warmed upto T~150 K. The measured data is plotted in Fig. 3 (blue squares) clearly shows that the cooling cycle ρ (T) is appreciably reduced, supercooling transition shifts to $T_{IM}^C \approx 85$ K but the warming IMT remains unchanged. The softening of the glassy state is evidenced by a decrease in ρ (5 K) by a factor of 1/3. The warming cycle ρ (T) gradually approaches the virgin value as $T_{IM}^W$ is reached. The most interesting feature is the disappearance of the $\rho_m(T_m)$ during the warming cycle. This clearly shows that the degree of freezing and randomness of the glassy state is drastically affected by the increase in the fraction of FMM phase, so much so that it has approached closer to the equilibrium. During the warming cycle ρ (T) shows a very weak temperature dependence up to T~50 K (ρ (56 K) = 1.035xρ(5 K)). At T>50 K, the ρ (T) rises steeply and gradually approaches the warming cycle IMT value. This clearly shows that the resistivity minimum in this case is a consequence of thermally activated devitrification of the glassy state and vanishes due to the softening of the glass at higher FMM fractions.

In order to study the impact of H on the AFM/COI to FMM phase transition, the supercooling transition, the frozen glassy state and $\rho_m(T_m)$ we measured ρ (T) in cooling and warming cycle at H= 10 kOe. As seen in Fig. 3 (navy diamonds), in the cooling cycle ρ (T) drops sharply and $T_{IM}^C$ jumps up from ≈56 K to ≈91 K. The drop in the low temperature resistivity ρ (5 K, H=10 kOe) by than a factor of 1/6 of the value at H=0 kOe clearly shows H induced melting of the magnetic glass. During warming $\rho_m(T_m)$ is conspicuously absent and ρ (T) first increases increase slowly ((ρ (56 K) = 1.015xρ(5 K)) and then rises sharply to peak at $T_{IM}^W$ (H=10 kOe) ≈ 150 K. Thus at H= 10 kOe the $T_{IM}^W$ is enhanced by 19 K and $\Delta T_{IM}$ is lowered from 75 K to 59 K. The data presented above clearly shows that, both the selective thermal recycling after cooling down from $T^*<T_{IM}^W$ and an external magnetic field (H = 10 kOe) reduce the degree of magnetic frustration in the liquid like state, lower the randomness in the magnetic glass state and hence lead to the vanishing of $\rho_m(T_m)$. However, thermal recycling only affects the non-equilibrium states, like the supercooled magnetic liquid, randomly frozen glassy state and $\rho_m(T_m)$, while the equilibrium state achieved during warming remains unaffected. In contrast, H has significant impact on non-equilibrium as well as equilibrium states of the system.

In manganites thin films structural/microstructural perturbations can also lead to the destruction or quenching of the AFM/COI phase with a concomitant enhancement in the FMM fraction. The possible way to create structural/microstructural perturbations is the substrate induced strain and its relaxation in single crystalline thin films. The structural characterization of the 4 hr annealed film discussed above has shown that this film is under compressive strain. In order to induce relaxation of the strain and hence increase the defect density we annealed one of the 4 hr annealed films for another 6 hrs at 850 °C in ~360 torr oxygen pressure. After annealing, the out of plane lattice constant of this film was reduced (from $a_c^{4H}$ 3.844 Å) to $a_c^{10H}$ = 3.825 Å, which almost coincides with that of the bulk. This clearly shows absence of any strain in this film and hence we conclude that annealing for longer duration results in complete relaxation of the strain in the present ~100 nm thick films. The FWHM of the (002) ω-scan of this film was found to be $\Delta\omega \approx 0.26°$ (inset of Fig. 4), which is about three times the value for the 4 hr annealed film ($\Delta\omega \approx 0.09°$). This broadening of the ω-scan is a clear signature of the enhanced defect density. The broadening of the ω-scan is generally caused by presence of dislocations, mosaic spread and curvature.[26] In the present case the Δω was found to be invariant with respect to the X-ray beam size and the peak position remained invariant of the spatial location of the beam on the film. These results confirm that the major source of the rocking curve broadening is the enhanced dislocation density. In manganites the cooperative J-T distortion that favours the AFM/COI phase is disrupted by the presence of dislocations.[1-5,23] This leads to carrier delocalization and enhancement FMM phase fraction. In view of this it is expected that the 10 hr annealed film has a higher FMM phase fraction and hence higher $T_{IM}$. To confirm this we measured the ρ-T of the 10 hr annealed thin film using the same protocol as used for the 4 hr annealed film. The ρ-T data of the 10 hr film is shown in Fig. 4. For comparison the ρ-T data of the 4 hr film has also been plotted. Although, the ρ-T in the PM regime remains nearly the same as that of the 4 hr annealed film, it is appreciable in the lower temperature region. The 10 hr annealed film shows $T_{IM}^C \approx 100$ K and $T_{IM}^W \approx 149$ K during cooling and warming cycles, respectively. It is interesting to note that ρ-T of the 10 hr annealed film almost resembles the ρ-T of the 4 hr film measured at 10 kOe. In addition to the enhanced IMT, the other noteworthy feature is the absence of $\rho_m(T_m)$. These results clearly confirm that increase in defect density leads to carrier delocalization and enhanced fraction of the FMM phase, which in turn affect the characteristic of the magnetic liquid and the glassy state by reducing magnetic frustration. This is well evidenced by the noticeable reduction in the hysteresis of the ρ-T curves and smaller value of $\Delta T_{IM}$=49 K. A dilution of the magnetic liquid character directly affects the glass like state in the lower temperature regime, which appears to be absent in the 10 hr annealed film.

In summary the origin of the robust resistivity minimum in a strongly phases separated manganite thin film has been studied as a function of the defect density, thermal cycling and magnetic field. Our results conclusively show that the resistivity minimum in strongly phase separated low bandwidth manganite, e.g., like $La_{1-x-y}Pr_yCa_xMnO_3$ is a consequence non-equilibrium states resulting from the magnetic frustration created by the delicate coexistence of the FMM and AFM/COI phases. The magnetic frustration first leads to the formation of a supercooled magnetic liquid, which when cooled down to further lower temperatures, freezes into a random glassy state. The thermally activated devitrification of non-equilibrium magnetic glass to equilibrium FMM phase lowers the resistivity and gives rise to the minimum. If the system is pushed towards the equilibrium by increasing the FMM phase fraction the minimum vanishes. Our study shows that this can be achieved by an external magnetic field, selective thermal cycling and increase in the defect density.

Authors are grateful to Prof. R. C. Budhani for his persistent encouragement. Dr. Anurag Gupta and Dr. V. P. S. Awana are thankfully acknowledged for magnetic (MPMS-DST facility) and magnetotransport measurement, respectively. Financial support from CSIR is thankfully acknowledged.

Figure Captions

Figure 1: 2θ-ω scan of the 4 hr annealed $La_{1-x-y}Pr_yCa_xMnO_3$ (x≈0.38, y≈0.42) thin film. The inset shows the rocking curve (ω scan) of the (002) peak.

Figure 2: Temperature dependence of dc magnetization (M-T) measured using ZFC, FCC and FCW protocols at H=100 Oe applied parallel to the film surface along the longer dimension. The inset shows the first order temperature derivative of magnetization.

Figure 3: Temperature dependence of resistivity (ρ-T) measured in cooling and warming cycles at H=0 kOe (black symbols) and H=10 kOe (red symbols). The ρ-T measured employing the selective recycling in which the sample was cooled down to 5 K and them warmed to 129 K and then again down to 5 K and warmed up to 150 K is plotted in blue symbols.

Figure 4: Temperature dependence of resistivity (ρ-T) of 4 hr (black symbols) and 10 hr (blue symbols) annealed film measured in cooling and warming cycles at H=0 kOe. The data of the 4 hr film measured at 10 kOe has also been included for comparison.